\documentclass[journal]{IEEEtran}
\usepackage{amsfonts} 
\usepackage{amsmath}
\usepackage{graphicx}

\usepackage[hyphens]{url}
\usepackage{xurl}            
\usepackage{hyperref}

\begin{document}
\title{A Spectral-Based Tuning Criterion for PI Controllers in IPDT Systems With Unified Tracking and Disturbance Rejection Performance
}

\author{Dhamdhawach~Horsuwan
\thanks{H. Dhamdhawach is with the School of Information Science and Technology, Vidyasirimedhi Institute of Science and Technology (VISTEC), Rayong 21210, Thailand (e-mail: dhamdhawach.h\_s23@vistec.ac.th).}}

\markboth{}{}

\maketitle

\begin{abstract}
This paper proposes a spectral-based tuning method for proportional–integral (PI) controllers in integrating-plus-dead-time (IPDT) systems. The design objective is to achieve unified exponential decay for both reference tracking and disturbance rejection by minimizing the spectral abscissa of the closed-loop system. A second-order semi-discrete model accurately captures the integrator and delay dynamics while enabling efficient dominant pole extraction. These discrete-time poles are mapped to continuous time and refined using Newton–Raphson iterations on the exact transcendental characteristic equation. The method produces a unique PI gain set without requiring heuristic trade-offs or weighting parameters. Comparative simulations demonstrate that the proposed tuning achieves faster convergence and improved robustness margins compared to classical rules (Ziegler–Nichols, SIMC) and integral performance criteria (IAE, ITAE). The approach provides a transparent and computationally efficient framework for PI control in delay-dominant systems.
\end{abstract}


\IEEEpeerreviewmaketitle
\section{Introduction}

Proportional--Integral (PI) controllers remain the predominant choice in industrial process control due to their simplicity, ease of implementation, and ability to achieve both reference tracking and disturbance rejection. As the minimal feedback configuration capable of eliminating steady-state error in regulating integrating or type-1 systems, PI controllers are widely employed across diverse industrial applications.

Many processes can be modeled as either First-Order Plus Dead Time (FOPDT) or Integrating Plus Dead Time (IPDT) systems. Under conditions of high integral gain or small process time constants, FOPDT dynamics often approximate those of an IPDT model. Consequently, the IPDT framework offers both generality and analytical convenience for analyzing delay-dominant systems.

Time delays are pervasive in practical control systems, arising from sensor dynamics, actuator backlash, communication latencies, or transport phenomena such as conveyor systems. These delays can significantly degrade control performance and, if not properly accounted for, may compromise closed-loop stability. Despite their practical importance, delays remain challenging to address rigorously within classical control design frameworks.

For IPDT systems under proportional-only control, the ultimate gain $K_u$ and oscillation period $P_u$ are directly related to the process parameters:
\[
K_u = \frac{\pi}{2LK}, \quad P_u = 4L,
\]
where $K$ is the process gain and $L$ is the time delay. These expressions form the basis for ultimate sensitivity tuning and process identification.

Multiple tuning strategies have been proposed for the IPDT configuration. Classical methods, such as Ziegler--Nichols (ZN)~\cite{ziegler1942optimum}, Cohen--Coon~\cite{cohen1953cohen}, and Chien--Hrones--Reswick (CHR)~\cite{CHR1952}, offer heuristic rules derived from step response characteristics or sustained oscillations. While straightforward to implement, these approaches often compromise performance optimality.

More systematic approaches, such as the SIMC method~\cite{skogestad2003simple}, yield explicit tuning formulas based on reduced-order models. However, these formulations typically rely on first-order dynamics and crude delay approximations, which may oversimplify the underlying system.

Frequency-domain techniques, including phase margin, gain margin, and bandwidth specifications, offer useful robustness metrics but do not directly quantify time-domain behaviors such as reference tracking and disturbance rejection. Likewise, graphical methods such as root-locus and pole-placement provide valuable design insights, but when applied to systems with transport delays, they almost universally require finite-order Pad\'e approximations~\cite{astrom1995pid}, which obscure the infinite-dimensional nature of delay dynamics and may produce misleading conclusions.

To address these limitations, numerical methods such as Chebyshev collocation~\cite{BREDA2006163} and semi-discretization~\cite{insperger2011semi} have been developed to discretize the time domain and approximate the system's infinite spectrum. These methods enable estimation of the dominant (rightmost) poles, facilitating stability and performance analysis for delay systems under a given controller.

Such numerical tools support continuous-time pole placement~\cite{wang2014} and PI tuning based on spectral abscissa optimization, which directly minimizes the real part of the slowest decaying closed-loop pole as a unified performance criterion~\cite{Optimal_Spectral2024, Pyragas2014}. This formulation naturally links controller design to the system's exponential decay characteristics.

In this paper, we extend this framework by formulating a PI controller design for IPDT systems that explicitly minimizes the spectral abscissa. Since the dominant pole governs the system's slowest convergence mode, the proposed method ensures that both reference tracking and disturbance rejection exhibit identical exponential decay rates.

Importantly, this approach eliminates the need for manually selected performance weights as required in IAE- or ITAE-based multi-objective designs. As shown in~\cite{performancs_tradeoff}, assigning weights between tracking and disturbance rejection often involves heuristic trade-offs. In contrast, the proposed spectral criterion provides a unified, dynamics-based performance measure that simultaneously addresses both objectives.

\section{System Model and Problem Formulation}

We consider an Integrating Plus Dead Time (IPDT) process, which serves as a representative model for many slow industrial systems, including temperature regulation, liquid-level control, and other transport-dominated processes. The process is described by the transfer function:
\begin{equation}
G(s) = \frac{K}{s} e^{-Ls},
\end{equation}
where $K > 0$ is the process gain and $L > 0$ is the time delay. The presence of the exponential delay term $e^{-Ls}$ introduces an infinite set of characteristic roots, which complicates both stability analysis and controller design~\cite{richard2003time}.

The controller under consideration is a standard proportional--integral (PI) compensator:
\begin{equation}
C(s) = K_P + \frac{K_I}{s},
\end{equation}
where $K_P$ and $K_I$ are the proportional and integral gains, respectively. Applying unity feedback yields the closed-loop characteristic equation:
\begin{equation} \label{eq:charactor}
1 + C(s)G(s) = 1 + \left(K_P + \frac{K_I}{s}\right)\frac{K}{s}e^{-Ls} = 0.
\end{equation}
Due to the presence of the delay term,~\eqref{eq:charactor} is transcendental and admits infinitely many roots in the complex plane. The closed-loop dynamics are primarily governed by the dominant roots, i.e., those with the largest real parts, which determine asymptotic stability and convergence rates.

Identifying these dominant poles generally requires numerical methods. Two prominent approaches are the Chebyshev collocation method~\cite{Optimal_Spectral2024, Pyragas2014}, implemented for example in MATLAB's BIFTOOL, and the semi-discretization method~\cite{insperger2011semi}.

In this work, we adopt the semi-discretization approach with second-order accuracy to approximate the delay system dynamics. The resulting finite-dimensional model provides initial estimates of the dominant poles. These estimates are subsequently refined using a Newton-based root-finding algorithm applied directly to the exact transcendental characteristic equation~\eqref{eq:charactor}, yielding high-precision localization of the dominant poles and enabling smooth continuation with respect to controller parameters.

\section{Discrete-Time Approximation and Spectral Analysis}

To efficiently explore admissible PI gain configurations and accurately predict dominant poles, we construct a reduced-order discrete-time approximation of the closed-loop IPDT system. In contrast to Pad\'e approximations or finite-dimensional lifted representations of delay systems~\cite{richard2003time}, this formulation retains the input delay explicitly and leverages the structure of the PI controller to minimize model dimensionality.

The input delay $L$ is discretized into $M$ equal segments of duration $h = L / M$, which we normalize as $h = 1/M$. The model tracks the current error $e[k]$ together with a memory buffer of the most recent $M+1$ control inputs $u[k], u[k-1], \ldots, u[k-M]$. The integral state $s[k]$ is not stored explicitly but is computed recursively using a fourth-order accurate numerical integration~\cite{semi-highorder}. This yields a compact state vector of dimension $M+2$:
\begin{equation}
\label{eq:state-vec}
x[k] = \begin{bmatrix} e[k] & u[k] & u[k-1] & \cdots & u[k-M] \end{bmatrix}^\top.
\end{equation}

The system dynamics follow from the IPDT relation $e'[k] = -K u[k - M]$, combined with the PI control law $u[k] = K_P e[k] + K_I s[k]$. The error is updated via a second-order finite-difference approximation:
\begin{equation}
e[k+1] = e[k] - \frac{hK}{2} \left( u[k - M] + u[k - M + 1] \right).
\end{equation}
The integral state is updated using fourth-order numerical integration:
\begin{align}
s[k+1] =\; & s[k] + \tfrac{h}{2} (e[k] + e[k+1]) \nonumber \\
          & - \tfrac{K h^2}{12} (u[k - M + 1] - u[k - M]).
\end{align}
Substituting into the control law yields:
\begin{equation}
u[k+1] = K_P e[k+1] + K_I s[k+1],
\end{equation}
which enables full state propagation from known quantities at time step $k$.

Unlike conventional discrete approximations such as zero-order hold (ZOH) models or first-order semi-discretizations~\cite{insperger2011semi}, this approach employs higher-order integration for both delay and integral terms~\cite{semi-highorder}, thereby enhancing model fidelity without relying on Pad\'e approximations or fully infinite-dimensional formulations. The result is a numerically efficient yet structurally precise model suitable for analyzing the closed-loop response under PI control.

While closely related to the semi-discretization framework of Insperger and Stépán~\cite{insperger2011semi, semi-highorder}, the present formulation differs significantly in both structure and objective. Classical semi-discretization lifts the system into a high-dimensional discrete-time representation for Floquet or monodromy analysis, typically employing first-order integration schemes. In contrast, the present model operates directly in discrete time, maintains a compact state space, and is specifically tailored for dominant pole extraction and time-domain simulation.

The semi-discrete approximation results in the linear recurrence:
\begin{equation}
x[k+1] = A(K_P,K_I)\,x[k],
\end{equation}
where $\{\lambda_i\}$ denote the eigenvalues of $A$.

\section{Continuous-Time Root Refinement for an IPDT Plant with PI Control}

To perform spectral tuning, we first extract the three dominant eigenvalues of the discrete-time system, denoted by $\{\lambda_i\}$, and map each back to continuous time via
\begin{equation}
\label{eq:si}
   \hat{s}_i = \frac{1}{h}\,\ln\bigl(\lambda_i\bigr),
   \quad i=1,2,3,
   \quad h=\frac{L}{M}.
\end{equation}
These preliminary estimates $\{\hat{s}_i\}$ serve as initial guesses for refinement using Newton--Raphson iterations applied to the exact transcendental characteristic equation of the closed-loop system:
\begin{equation}
\Delta(s) = s^{2} + K\left(K_{P}\,s + K_{I}\right)\,e^{-L s} = 0,
\label{eq:ipdt_pi_char}
\end{equation}
where $K$ is the process gain, $L$ is the time delay, and $K_P$, $K_I$ are the controller gains. For each initial guess $\hat{s}_i$, the refinement procedure yields one real root and one complex-conjugate pair, producing the dominant pole set $\{s_i\}$ that governs closed-loop behavior.

The control design objective is to minimize the slowest decay rate among these dominant poles, defined by the spectral abscissa
\begin{equation}
J(K_P, K_I) = \max_{i}\Re\{s_i\}.
\end{equation}
Because $J$ is generally non-smooth and non-convex over the $(K_P, K_I)$ plane, we solve the optimization problem
\begin{equation}
K^\star_P, K^\star_I = \arg\min_{K_P,K_I} J(K_P,K_I),
\end{equation}
using the differential evolution algorithm of Storn and Price~\cite{storn1997differential}, as implemented in \texttt{SciPy}~\cite{scipy_optimize_de}. This global optimization framework efficiently navigates the nonconvex gain space and consistently identifies a unique minimizer corresponding to the global spectral valley.

An important feature of the resulting optimum is the nearly vertical alignment of the dominant poles in the complex-$s$ plane: the real pole and the complex-conjugate pair exhibit approximately identical real parts. This spectral symmetry ensures that both oscillatory and non-oscillatory components of the response decay at similar rates, yielding uniform transient behavior across both reference tracking and disturbance rejection scenarios.

\begin{figure}[!t]
    \centering
    \includegraphics[width=\linewidth]{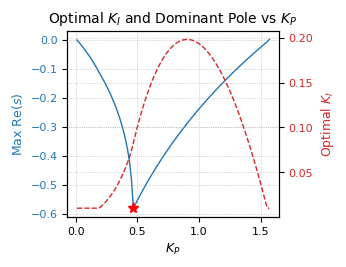}
    \caption{Unique optimal $K_I$ values corresponding to each $K_P$ value. The blue curve shows the minimum real part of the dominant poles (cost), while the red dashed curve traces the optimal $K_I$. Randomized $K_I$ samples per $K_P$ further confirm the uniqueness and smoothness of the spectral valley.}
    \label{fig:optimal_KI_sweep}
\end{figure}

\begin{figure}[!t]
    \centering
    \includegraphics[width=\linewidth]{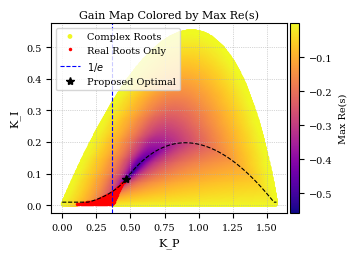}
    \caption{Stability region in the $(K_P, K_I)$ plane. The background contour represents the maximum real part of dominant poles obtained from continuous-time refinement. The black curve traces the optimal $K_I$ for each $K_P$. The black star marks the global optimum. The red region indicates configurations with a real dominant root.}
    \label{fig:stability_region_optimal}
\end{figure}

\begin{figure}[!t]
    \centering
    \includegraphics[width=\linewidth]{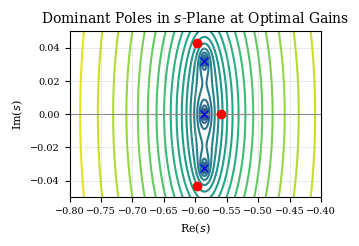}
    \caption{Dominant poles (blue crosses) on the complex-$s$ plane at the optimal PI gains. Initial seeds from the second-order discrete model with $M = 20$ are marked in red. The background contour represents $\log_{10}|F(s)|$, where $F(s)$ is the closed-loop characteristic function.}
    \label{fig:optimal-poles}
\end{figure}

\section{Trajectory Tracking and Disturbance Rejection}

The time-domain performance is evaluated for both reference tracking and disturbance rejection using a unit-step reference input and a step disturbance at the control input, both formulated within the same semi-discrete model framework.

\paragraph{Reference tracking:}  
For a unit-step reference input $r[k] = 1$ (in the absence of disturbance), the initial conditions are set as
\[
e[0] = 1, \quad s[0] = 0, \quad u[0] = K_P e[0] + K_I s[0] = K_P,
\]
which yields the initial state vector
\[
x[0] = \begin{bmatrix} 1 & K_P & 0 & \cdots & 0 \end{bmatrix}^\top.
\]

\paragraph{Disturbance rejection:}  
For a constant disturbance $d[k] = D$ entering additively at the controller output, the PI control law becomes
\[
u[k] = K_P e[k] + K_I s[k] + D.
\]
Introducing a shifted integrator state
\[
\tilde{s}[k] = s[k] + \frac{D}{K_I},
\]
the control law may be rewritten as
\[
u[k] = K_P e[k] + K_I \tilde{s}[k].
\]
Under a step disturbance applied at zero reference input ($r[k] = 0$), the initial conditions become $e[0] = 0$ and $\tilde{s}[0] = D/K_I$, yielding
\[
u[0] = D, \quad x[0] = \begin{bmatrix} 0 & D & 0 & \cdots & 0 \end{bmatrix}^\top.
\]
Thus, both reference tracking and disturbance rejection are represented as distinct initial conditions within the same semi-discrete state space model.

\paragraph{Unified eigenstructure interpretation:}  
In this formulation, reference tracking primarily excites the complex-conjugate pair of dominant poles, while disturbance rejection—mediated by integral action—primarily excites the real pole. When the dominant poles are aligned vertically in the complex plane, all modes exhibit identical exponential decay rates, ensuring uniform convergence behavior for both reference tracking and disturbance rejection. This structure yields a unified and worst-case-optimal transient response across both scenarios.

\section{Comparison with Classical PI Tuning Methods}

To assess the effectiveness of the proposed tuning criterion, its performance is compared against two widely used PI tuning rules for delay-dominant systems: the Ziegler--Nichols (ZN) ultimate sensitivity method~\cite{ziegler1942optimum} and the Skogestad Internal Model Control (SIMC) rule~\cite{skogestad2003simple}. Both aggressive and conservative SIMC variants are considered.

All methods are evaluated on the same normalized IPDT system with transfer function $G(s) = \frac{1}{s} e^{-s}$. Controllers are implemented in continuous time, and performance is assessed based on step responses for both reference tracking and disturbance rejection.

\begin{table}[!t]
\centering
\caption{PI Gains from Classical and Proposed Methods}
\begin{tabular}{lcc}
\hline
\textbf{Strategy} & $K_P$ & $K_I$ \\
\hline
Proposed Method & 0.4614 & 0.0793 \\
Ziegler--Nichols & 0.7069 & 0.2121 \\
SIMC (Conservative) & 0.2857 & 0.0204 \\
SIMC (Aggressive) & 0.9524 & 0.2268 \\
\hline
\end{tabular}
\label{tab:pi-gains}
\end{table}

\begin{figure}[!t]
    \centering
    \includegraphics[width=\linewidth]{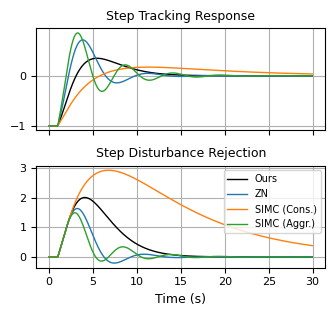}
    \caption{Step responses under various PI tuning methods: Ziegler--Nichols, SIMC (aggressive and conservative), and the proposed approach.}
    \label{fig:step_classic}
\end{figure}

The step responses under each tuning method are shown in Fig.~\ref{fig:step_classic}. The proposed method achieves faster convergence than the conservative SIMC tuning and provides reference-tracking performance comparable to the ZN method, while significantly reducing oscillations. In disturbance rejection, the proposed controller exhibits slightly higher overshoot than ZN and aggressive SIMC but maintains a fully non-oscillatory response profile.

\begin{figure}[!t]
    \centering
    \includegraphics[width=\linewidth]{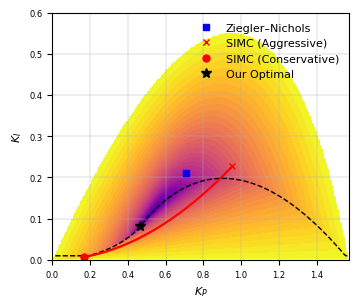}
    \caption{Gain trajectories of classical methods (ZN, SIMC) overlaid on the dominant pole contour and optimal $K_I$ curve.}
    \label{fig:classic_region}
\end{figure}

Figure~\ref{fig:classic_region} depicts the gain trajectories of the classical methods overlaid on the dominant pole contour map together with the optimal $K_I$ boundary derived from the proposed spectral approach. It is observed that the SIMC gain trajectories closely follow the optimal $K_I$ boundary, while the ZN tuning point lies in proximity to this optimal region. This alignment suggests that, despite being derived from different principles, these classical methods partially coincide with the numerically optimal region identified by the proposed unified criterion.

\section{Comparison with Optimization-Based Methods}

While classical tuning rules offer simplicity and interpretability, modern control design often employs numerical optimization of performance indices. To further evaluate how the proposed spectral criterion aligns with such approaches, we compare it with PI parameters obtained by minimizing integral performance indices, specifically the Integral of Absolute Error (IAE) and the Integral of Time-weighted Absolute Error (ITAE).

\begin{figure}[!t]
    \centering
    \includegraphics[width=\linewidth]{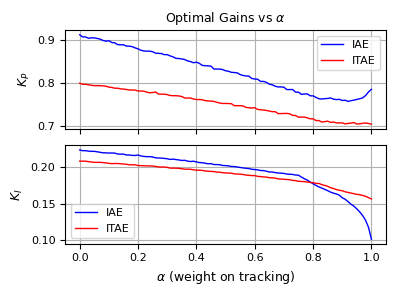}
    \caption{PI gain trajectories for IAE and ITAE as a function of weighting parameter $\alpha$.}
    \label{fig:opt-gain}
\end{figure}

Each method introduces a weighting parameter $\alpha \in [0, 1]$ to balance the trade-off between reference tracking and disturbance rejection, where $\alpha = 1$ corresponds to pure trajectory tracking and $\alpha = 0$ corresponds to pure disturbance rejection. The resulting gain trajectories as functions of $\alpha$ are shown in Fig.~\ref{fig:opt-gain}.

For quantitative comparison, we select $\alpha = 0.5$, representing an equal weighting of both objectives. The corresponding PI gains are summarized in Table~\ref{tab:opt-gains}.

\begin{table}[!t]
\centering
\caption{PI Gains from Optimization-Based Methods ($\alpha = 0.5$)}
\begin{tabular}{lcc}
\hline
\textbf{Strategy} & $K_P$ & $K_I$ \\
\hline
Proposed Optimal & 0.4614 & 0.0793 \\
IAE ($\alpha = 0.5$) & 0.8289 & 0.2015 \\
ITAE ($\alpha = 0.5$) & 0.7532 & 0.1916 \\
\hline
\end{tabular}
\label{tab:opt-gains}
\end{table}

\begin{figure}[!t]
    \centering
    \includegraphics[width=\linewidth]{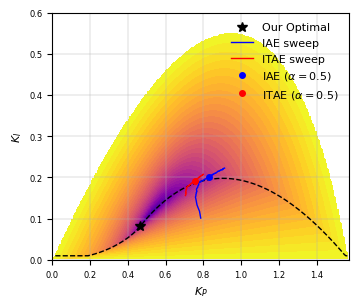}
    \caption{Gain regions of IAE and ITAE methods overlaid on the optimal $K_I$ contour.}
    \label{fig:opt-region}
\end{figure}

Figure~\ref{fig:opt-region} illustrates that the gain trajectories for both IAE and ITAE intersect the optimal $K_I$ boundary obtained from the proposed spectral approach. This intersection reflects the underlying trade-off between disturbance rejection and reference tracking: gain pairs located above the optimal boundary favor stronger integral action and improved disturbance rejection, whereas those below emphasize reference-tracking performance.

\begin{figure}[!t]
    \centering
    \includegraphics[width=\linewidth]{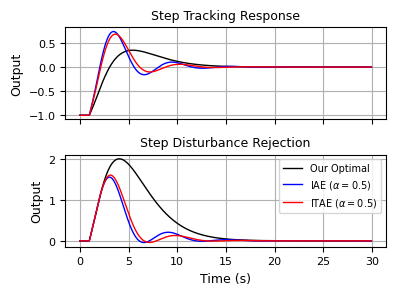}
    \caption{Step responses for trajectory tracking and disturbance rejection using IAE, ITAE ($\alpha = 0.5$), and the proposed method.}
    \label{fig:opt-step}
\end{figure}

The corresponding step responses for $\alpha = 0.5$ are shown in Fig.~\ref{fig:opt-step}. For reference tracking, the proposed method exhibits similar convergence speed but achieves reduced overshoot and complete suppression of oscillatory behavior compared to the IAE and ITAE designs. For disturbance rejection, while the proposed method produces slightly slower convergence and moderately higher peak overshoot, it maintains a fully non-oscillatory profile, in contrast to the more oscillatory transient behavior observed under IAE and ITAE tunings.

\section{Phase Margin and Gain Margin}

Using the optimal PI gains $(K_P^\star, K_I^\star)$ determined by the spectral-valley criterion, the closed-loop robustness is evaluated through classical stability margins. The loop transfer function is given by
\begin{equation}
L(j\omega) = K\,\frac{K_P^\star\,j\omega + K_I^\star}{(j\omega)^2}\,e^{-j\omega L}.
\end{equation}

\begin{figure}[!t]
  \centering
  \includegraphics[width=\columnwidth]{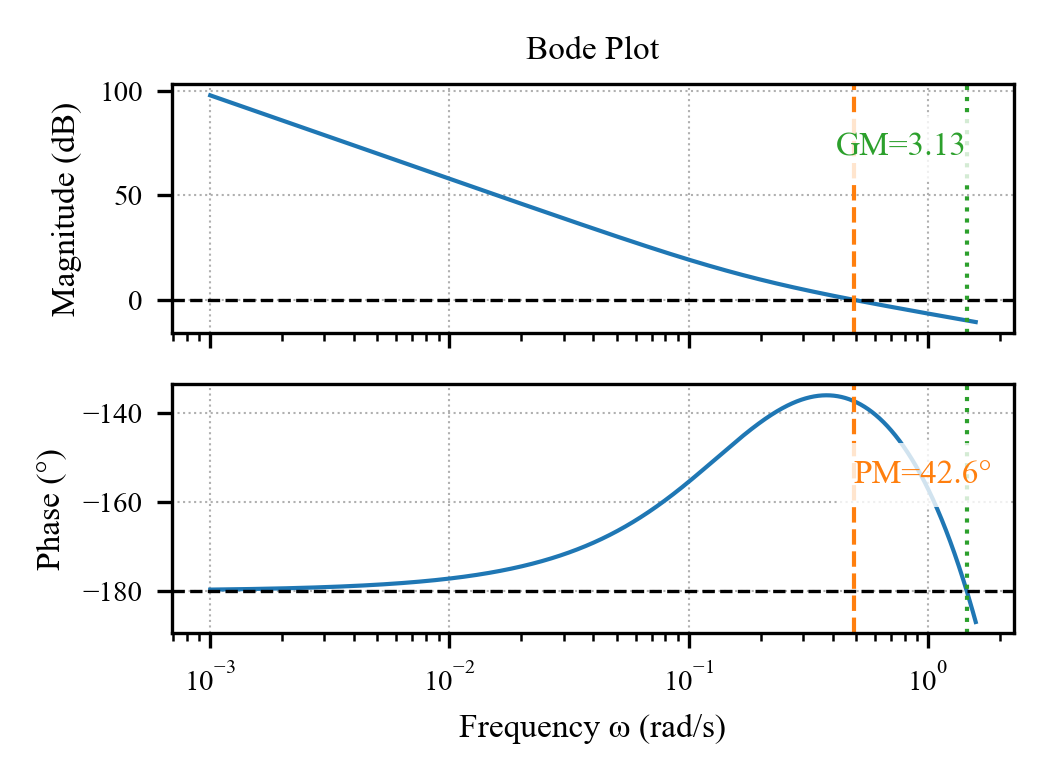}
  \caption{Bode magnitude and phase for the optimally tuned PI+IPDT loop with $K_P^\star=0.4614$, $K_I^\star=0.0793$, and $L=1$. The dashed horizontal lines indicate the gain- and phase-crossover levels. Vertical markers denote the gain-crossover at $\omega_{gc}\approx0.4891\,\mathrm{rad/s}$ (orange), yielding $\mathrm{PM}\approx42.6^\circ$, and the phase-crossover at $\omega_{pc}\approx1.4531\,\mathrm{rad/s}$ (green), yielding $\mathrm{GM}\approx3.13$ (9.9\,dB).}
  \label{fig:bode_optimal}
\end{figure}

The corresponding Bode magnitude and phase plots are shown in Fig.~\ref{fig:bode_optimal}. The dashed horizontal lines indicate the 0\,dB gain-crossover level and $-180^\circ$ phase-crossover level. The gain-crossover frequency and phase margin are computed as
\[
\omega_{gc} \approx 0.4891\,\mathrm{rad/s}, \quad \mathrm{PM} \approx 42.6^\circ,
\]
and the phase-crossover frequency and gain margin as
\[
\omega_{pc} \approx 1.4531\,\mathrm{rad/s}, \quad \mathrm{GM} \approx 3.13\;(9.9\,\mathrm{dB}).
\]

\begin{figure}[!t]
  \centering
  \includegraphics[width=\columnwidth]{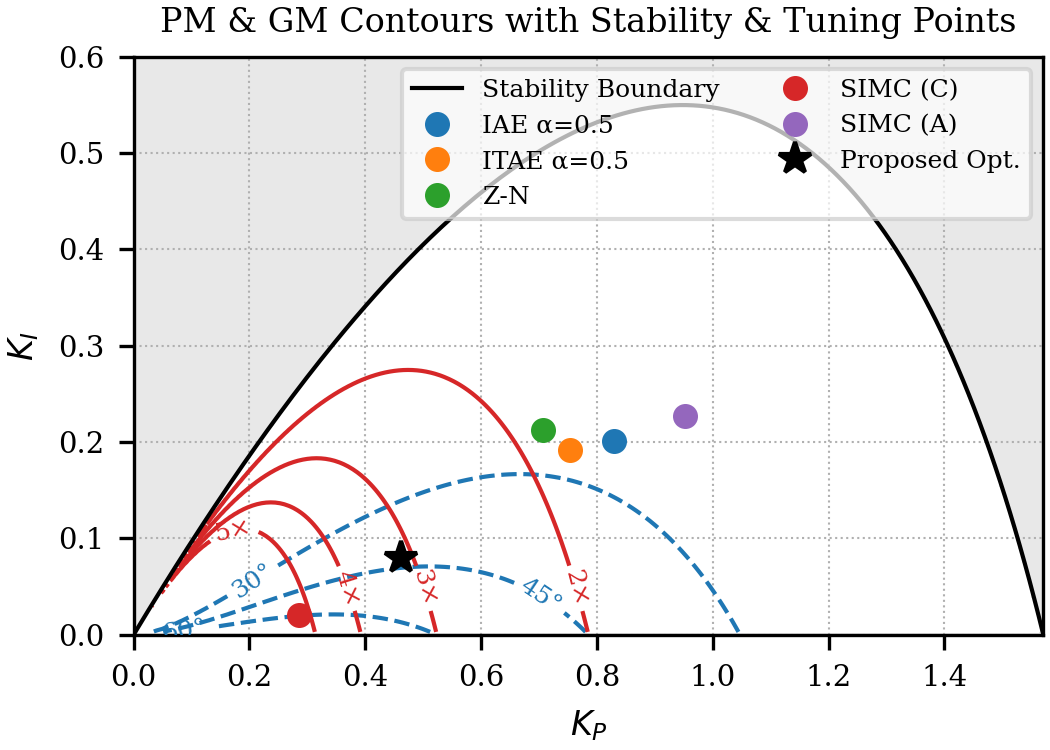}
  \caption{Contours of closed-loop phase margin (PM) and gain margin (GM) over the $(K_P,K_I)$ plane, with the stability boundary and classical tuning methods overlaid. Blue dashed curves: PM = 30°, 45°, 60°. Red solid curves: GM = 1×–5×. Light-gray region: unstable. Circles: classical tuning methods (IAE, ITAE, Ziegler–Nichols, SIMC conservative/aggressive). Star: proposed optimum $(K_P^\star,K_I^\star)$.}
  \label{fig:pm_gm_contours}
\end{figure}

To contextualize the proposed design within the broader $(K_P,K_I)$ parameter space, Fig.~\ref{fig:pm_gm_contours} presents phase and gain margin contours. The dashed blue curves correspond to phase margins of 30°, 45°, and 60°, while the solid red curves correspond to gain margins of 1×, 2×, 3×, 4×, and 5×. The light-gray region indicates instability, bounded by the Nyquist-derived stability limit (black curve). Gain pairs resulting from six classical tuning methods are also displayed, with the proposed optimum marked by a black star.

It is observed from Fig.~\ref{fig:pm_gm_contours} that the proposed gains reside well within the robust region, characterized by phase margins exceeding $40^\circ$ and gain margins exceeding $3$. The conservative SIMC tuning yields slightly larger margins but at the cost of significantly slower closed-loop response. The proposed method thus achieves a favorable compromise between robustness and transient speed, compared to more aggressive step-response-based rules that prioritize performance at the expense of stability margins.

\section*{Conclusion}

This paper has presented a novel dynamics-based tuning criterion for PI controllers in integrating-plus-dead-time (IPDT) systems, formulated through direct minimization of the spectral abscissa—the real part of the slowest closed-loop pole. By combining a high-fidelity semi-discrete approximation of the input delay with a continuous-time root-refinement procedure, the method identifies a unique gain pair $(K_P^\star,K_I^\star)$ that enforces symmetric exponential decay across both reference tracking and disturbance rejection tasks.

The proposed approach offers several key advantages:
\begin{enumerate}
    \item \textit{Unified performance metric:} Unlike multi-objective IAE/ITAE formulations or heuristic rules, the spectral abscissa encapsulates both transient speed and robustness in a single, physically interpretable cost function.
    \item \textit{Predictable robustness:} Classical Bode and gain/phase margin analyses confirm that the resulting design resides well within the stable region ($\mathrm{PM}\approx42.6^\circ$, $\mathrm{GM}\approx3.13$), second only to highly conservative tunings that yield substantially slower responses.
    \item \textit{Transparent trade-off visualization:} Contour maps in the $(K_P,K_I)$ plane provide clear geometric insight into how various tuning rules position themselves relative to the Nyquist stability boundary and robustness margins.
    \item \textit{Numerical efficiency and simulation fidelity:} The two-step procedure—consisting of rapid eigenvalue evaluation from a compact discrete-time model, followed by local Newton refinement—achieves high computational efficiency and accuracy. The discrete model itself may also serve as a high-fidelity platform for time-domain simulations.
\end{enumerate}

Comparative simulations demonstrate that the proposed design matches or exceeds the reference-tracking performance of aggressive tuning rules (e.g., Ziegler--Nichols, Aggressive SIMC, IAE, ITAE), while fully suppressing oscillatory behavior. In addition, it significantly outperforms conservative SIMC tunings in terms of settling time, thereby achieving a balanced compromise between transient speed, damping, and robustness.

Future work will focus on extending the spectral tuning framework to PID structures and more general industrial models, such as first-order plus dead-time (FOPDT) processes, as well as investigating adaptive spectral tuning strategies for real-time implementation under time-varying process dynamics.

\appendices

\section{Applying the Normalized Gains to Real-World Processes}

All analysis throughout the paper, and the optimal gains $(K_P^\star, K_I^\star) = (0.4614, 0.0793)$, are based on the normalized plant
\[
G_{\mathrm{norm}}(s) = \frac{1}{s} e^{-s},
\]
corresponding to a process gain $K=1$ and delay $L=1$. For a general IPDT process
\[
G(s) = \frac{K}{s} e^{-Ls},
\]
the normalized gains are scaled to actual controller parameters as
\[
K_{P,\mathrm{real}} = \frac{K_P^\star}{K L}, \qquad K_{I,\mathrm{real}} = \frac{K_I^\star}{K L^2}.
\]
The corresponding PI controller is thus implemented as
\[
C(s) = K_{P,\mathrm{real}} + \frac{K_{I,\mathrm{real}}}{s},
\]
where
\[
K_{P,\mathrm{real}} = \frac{0.4614}{K L}, \qquad K_{I,\mathrm{real}} = \frac{0.0793}{K L^2}.
\]
This simple scaling preserves the spectral-abscissa optimum and guarantees identical exponential decay rates for both tracking and disturbance rejection, independent of the specific process parameters $K$ and $L$.

\section{Discrete Model Evaluation}

The accuracy of the semi-discrete model is evaluated by comparing the dominant pole obtained from the continuous-time characteristic equation with that predicted by the discrete-time approximation, for a fixed delay resolution $M = 20$. The absolute real-part error is computed as
\begin{equation}
\Delta_{\max}(K_P,K_I) = \left|\Re\{s_{\mathrm{cont}}\} - \Re\{s_{\mathrm{approx}}\}\right|.
\end{equation}

\begin{figure}[!t]
    \centering
    \includegraphics[width=\linewidth]{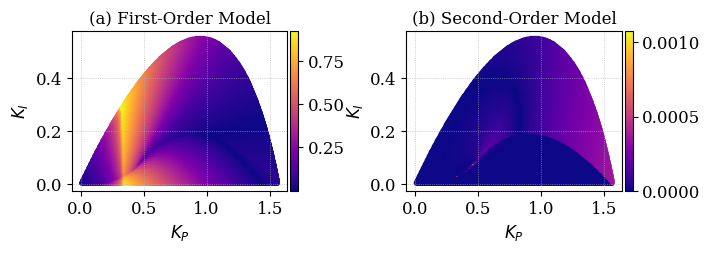}
    \caption{Absolute error in the real part of the dominant pole between the continuous-time root and discrete-time approximation at $M=20$. (a) First-order model. (b) Second-order model. Note differing colorbar scales: the second-order model achieves substantially lower error.}
    \label{fig:model_error_comparison}
\end{figure}

The resulting error distribution over the $(K_P, K_I)$ space is shown in Fig.~\ref{fig:model_error_comparison}. The first-order model [Fig.~\ref{fig:model_error_comparison}(a)] exhibits significant error, with deviations exceeding 0.8 for regions of high integral gain. In contrast, the second-order model [Fig.~\ref{fig:model_error_comparison}(b)] demonstrates substantially improved accuracy, with errors remaining below $10^{-3}$ across most of the stabilizing region. These results demonstrate that higher-order approximations are essential for accurately resolving input-delay dynamics.

\section{Comparison with Pad\'e Approximations}

\begin{figure}[!t]
  \centering
  \includegraphics[width=\columnwidth]{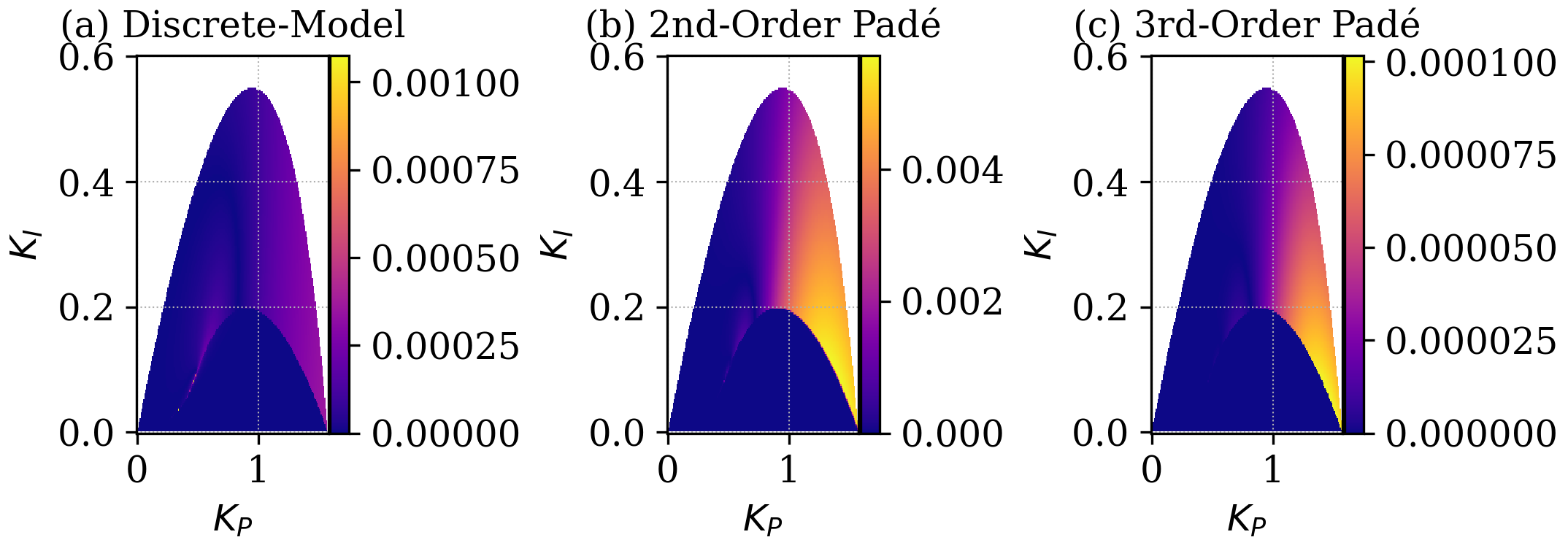}
  \caption{Error in the dominant continuous-time pole real part across $(K_P,K_I)$, shown only inside the stability region, for: (a) semi-discrete model ($M=20$), (b) second-order Pad\'e, and (c) third-order Pad\'e. Grey regions denote unstable gain pairs.}
  \label{fig:compare_pade}
\end{figure}

The approximation accuracy is further compared against Pad\'e models by evaluating, at each $(K_P,K_I)$, the absolute error in the real part of the dominant continuous-time pole. The error distributions for three methods are shown in Fig.~\ref{fig:compare_pade}, limited to the closed-loop stability region:  
(a) the proposed semi-discrete model with $M=20$,  
(b) the second-order Pad\'e approximation, and  
(c) the third-order Pad\'e approximation. 

Within the stability region, the semi-discrete approach yields lower error than second-order Pad\'e, while third-order Pad\'e achieves the smallest overall error.

\begin{figure}[!t]
  \centering
  \includegraphics[width=\linewidth]{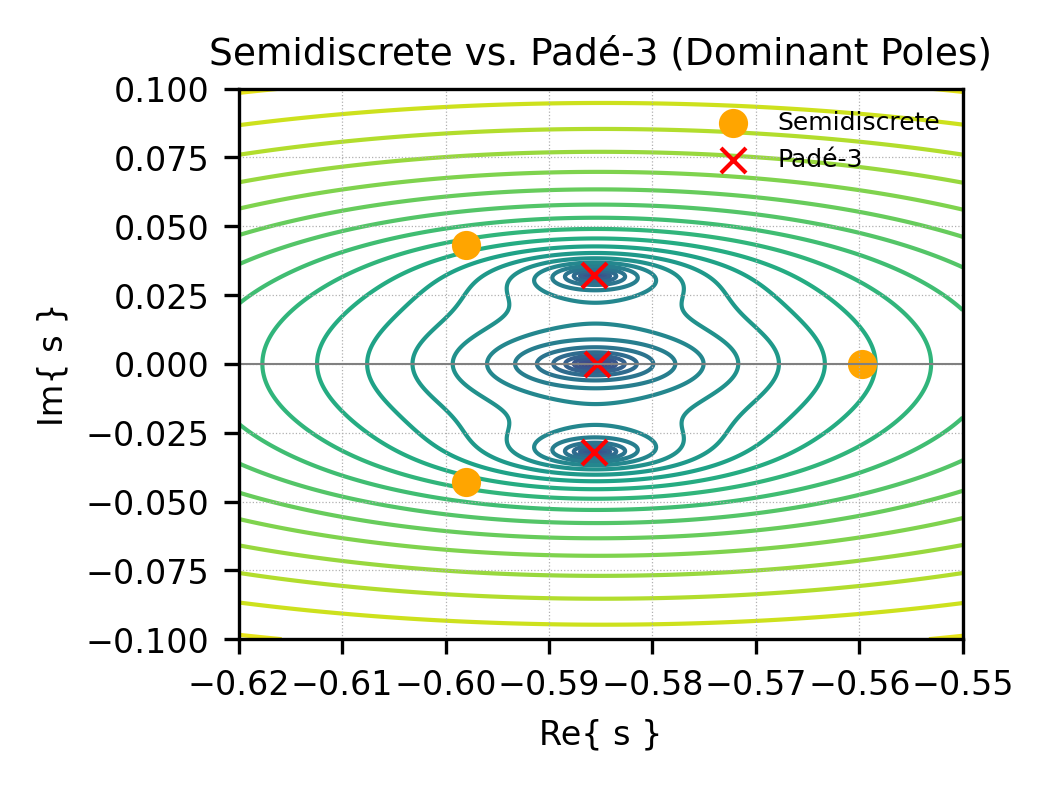}
  \caption{Zoomed comparison of semi-discrete and Pad\'e-3 poles within the dominant-pole cluster ($\Re\{s\} \in [-0.62, -0.55]$, $\Im\{s\} \in [-0.1, 0.1]$). Orange circles: semi-discrete; red crosses: Pad\'e-3.}
  \label{fig:dom_pole}
\end{figure}

\begin{figure}[!t]
  \centering
  \includegraphics[width=\linewidth]{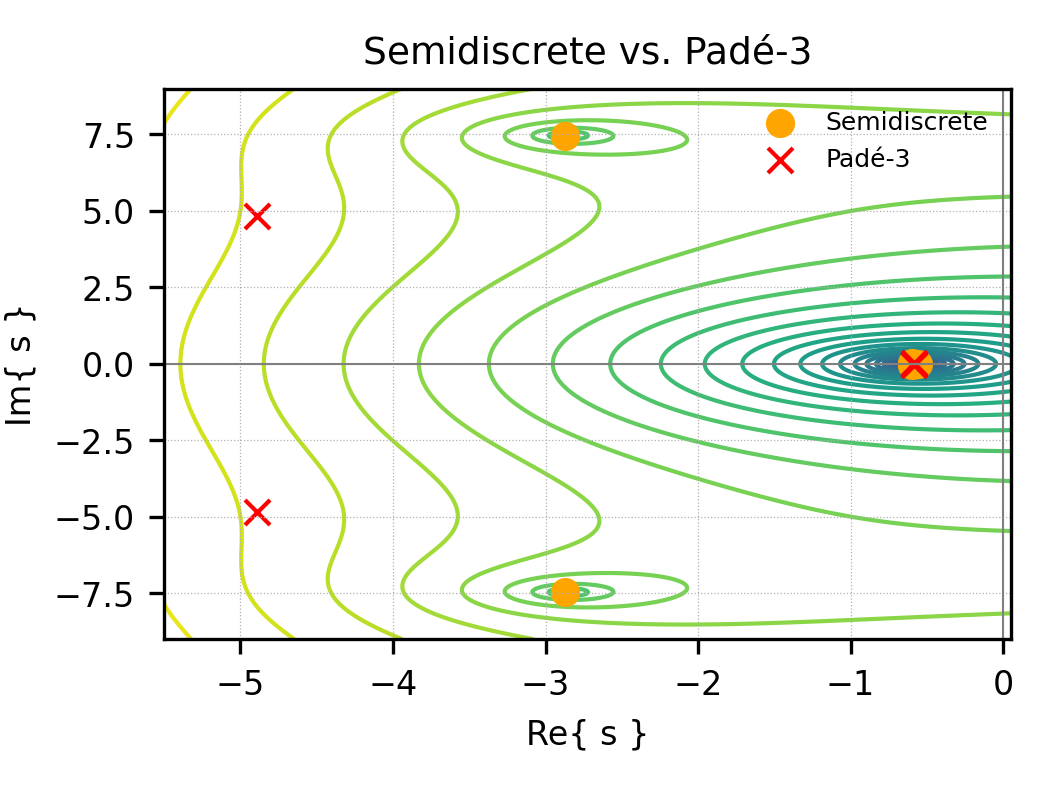}
  \caption{Full-scale contour of $\log_{10}\bigl|s^2 + K K_P s e^{-Ls} + K K_I e^{-Ls}\bigr|$, overlaid with pole seeds. Orange circles: semi-discrete poles; red crosses: Pad\'e-3 poles.}
  \label{fig:other_pole}
\end{figure}

For additional insight, Figs.~\ref{fig:dom_pole}--\ref{fig:other_pole} compare individual pole locations between the semi-discrete and Pad\'e-3 models. The zoomed view in Fig.~\ref{fig:dom_pole} focuses on the dominant pole cluster, showing that the Pad\'e-3 model closely matches the semi-discrete poles for these leading eigenvalues.

The full complex plane comparison in Fig.~\ref{fig:other_pole} further reveals that the Pad\'e-3 model deviates more significantly for non-dominant delay-induced poles, while the semi-discrete model better captures the full delay spectrum.

These results confirm that while the third-order Pad\'e approximation provides highly accurate estimates of the dominant poles for design and frequency-domain analysis, the proposed semi-discrete model offers superior fidelity for full-spectrum and time-domain simulation of delay-dominant systems.

\ifCLASSOPTIONcaptionsoff
  \newpage
\fi

\bibliographystyle{IEEEtran}
\bibliography{refs}

\begin{IEEEbiography}[{\includegraphics[width=1in,height=1.25in,clip,keepaspectratio]{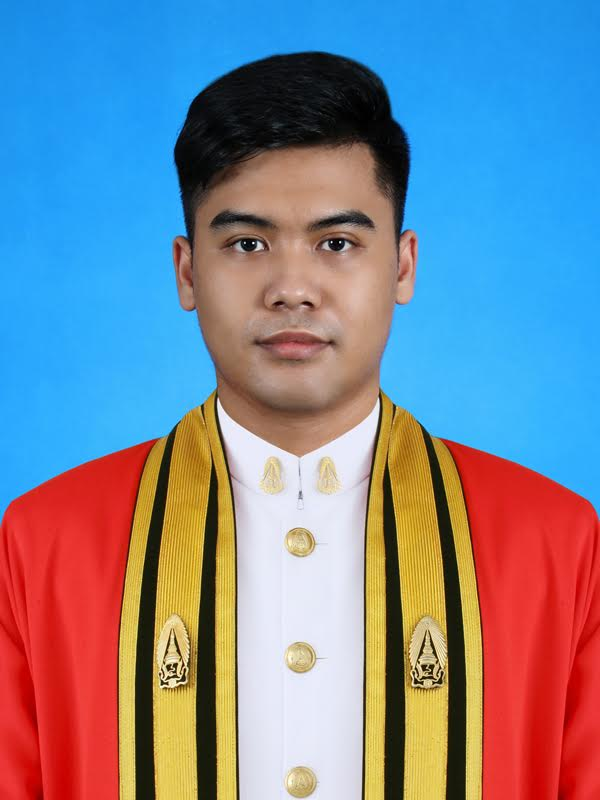}}]{Dhamdhawach Horsuwan}
received the B.Eng. degree in robotics and automation engineering from the Institute of Field Robotics (FIBO), King Mongkut's University of Technology Thonburi, Bangkok, Thailand, in 2020. He is currently pursuing the M.Eng. degree in information science and technology at the Vidyasirimedhi Institute of Science and Technology (VISTEC), Rayong, Thailand. His research interests include control systems, multi-agent coordination, and bio-inspired robotics.
\end{IEEEbiography}

\end{document}